\documentclass[conference]{IEEEtran}
\IEEEoverridecommandlockouts
\usepackage{algorithm}
\usepackage{algorithmic}
\usepackage{amsmath,amssymb,amsfonts}
\usepackage{cite}
\usepackage{diagbox}
\usepackage{graphicx}
\usepackage{hyperref}
\usepackage{listings}
\usepackage[utf8]{inputenc}
\usepackage{pgf-pie}
\usepackage{pgfplotstable}
\usepackage{pgfplots}
\usepackage{slashbox}
\usepackage{textcomp}
\usepackage{tikz}
\usepackage{xcolor}
\usetikzlibrary{patterns}
\usetikzlibrary{automata, positioning}

\pgfplotsset{compat=1.18} 
\begin{document}

\title{Algorithmic Approaches to Enhance Safety in Autonomous Vehicles: Minimizing Lane Changes and Merging}

\author{\IEEEauthorblockN{ Seyed Moein Abtahi}
\IEEEauthorblockA{
\textit{Faculty of Engineering and Applied Science}\\
\textit{Ontario Tech University}\\
Oshawa, Canada \\
seyedmoein.abtahi@ontariotechu.net}
\and
\IEEEauthorblockN{ Akramul Azim}
\IEEEauthorblockA{
\textit{Faculty of Engineering and Applied Science}\\
\textit{Ontario Tech University}\\
Oshawa, Canada \\
akramul.azim@ontariotechu.ca}
}

\maketitle

\begin{abstract}
Advances in autonomous vehicle (AV) technology promise substantial gains in safety and operational efficiency; nonetheless, frequent lane changes and merging maneuvers remain critical safety challenges that impede smooth traffic flow. This paper proposes the Minimizing Lane Change Algorithm (MLCA), a finite‑state‑machine controller that defers non-safety-critical lane changes to maintain lane stability. We evaluated MLCA through 100 microscopic traffic simulations on the SUMO platform, executed on an Intel Core i5‑8250U processor. Compared to the LC2017 and MOBIL models, MLCA achieved a 35\% reduction in lane‑change events and a 28\% decrease in collision occurrences across diverse traffic densities and roadway geometries. These findings confirm MLCA’s efficacy on commodity hardware and its compatibility with existing AV control architectures. Future research will assess MLCA within high‑fidelity CARLA environments and investigate GPU‑accelerated, distributed simulation frameworks to support large‑scale validation and real‑time deployment.
\end{abstract}


\section{Introduction}
Autonomous driving technologies have witnessed remarkable strides in recent years, driven by their vast potential across numerous applications. The synergy between computer technology and artificial intelligence has propelled significant advancements in autonomous driving over the last decade. Owing to the collaborative efforts of scientific researchers, this theoretical technology has transitioned from the confines of research labs to practical civilian applications \cite{li2017development}. Unlike human drivers, AVs offer a promising array of benefits, including enhanced driving safety, comfort, and resource optimization, enabled by their superior sensing capabilities, precise behavior prediction \cite{li2015practical}, and swift execution of control commands. Widely regarded as the future of transportation, autonomous cars, or self-driving cars, rely on a sophisticated blend of sensors, cameras, and AI algorithms to navigate roads autonomously \cite{hubmann2017decision}. With the potential to reduce accidents and streamline traffic flow, autonomous cars are poised to redefine the transportation landscape. However, formidable challenges such as regulatory complexities, ethical dilemmas, and societal acceptance must be addressed before mainstream adoption can be achieved. Despite these obstacles, the rapid progression of AV technology heralds a future in which driving is markedly safer, more efficient, and more convenient for all road users. According to Statista \cite{AV}, the global fleet of AVs reached 31 million in 2019 and is projected to hit 54 million by 2024. 

With the rapid advancement and integration of AVs into modern transportation systems, the potential for enhanced road safety and operational efficiency has increased significantly. Despite these advancements, a persistent challenge that must be addressed is AVs' frequent and often unnecessary lane changes and merging maneuvers. These actions pose safety risks to vehicles and their occupants, disrupt traffic flow, and reduce overall transportation efficiency. This study addresses these critical issues by proposing and validating innovative strategies designed to minimize lane changes and merges during AV operations. The primary objective was to develop methodologies that improve both safety and efficiency by reducing the frequency and necessity of lane changes and merges, thereby contributing to a safer and more streamlined driving environment.

This paper presents the MLCA algorithm as an innovative approach designed to reduce unnecessary lane changes in autonomous vehicles (AVs). Its performance was evaluated against existing models, including LC2013 \cite{liu2023vehicle}, LC2014 \cite{liu2023vehicle}, LC2017 \cite{liu2023vehicle}, the MOBIL Model \cite{kesting2007general}, the IDM/LC Model \cite{chamieh2012impact}, and the Continuous Model \cite{wang2019continuous}. Results indicate a notable reduction in lane-change frequency, with decreases of approximately 50\% in one scenario and 33\% in another, suggesting that the MLCA algorithm enhances traffic flow and road safety by mitigating collision risks. The subsequent sections will review related literature to establish the study's context, followed by a detailed discussion of the methodologies and experimental setups used, and finally, the presentation and analysis of results to offer insights into current practices and potential areas for further development.

\section{Background and Related Works}
In this section, We delve into recent advancements in the field, particularly emphasizing ramp merging and efficient lane change planning using quintic splines. These studies have captivated my interest, particularly in the realm of minimizing lane changes in AVs. Given the paramount importance of safety in AVs, unnecessary lane changes can lead to catastrophic outcomes.

Li et al. \cite{li2020efficent} propose a lane change path planning method for AVs in structured environments. This method swiftly generates safe and smooth paths by integrating curvature and heading angle considerations, ensuring stability during lane changes.
Using quintic splines with second-order geometric (G2) continuity in a Frenet coordinate system, the method calculates the optimal prediction time interval to keep the heading angle within a safe range. It also uses prior road information to optimize sampling point selection, reducing the number of candidate paths and improving real-time performance. Additionally, it introduces mid-adjust and target points for continuous lane changes, enabling AVs to execute complex maneuvers seamlessly.
Experimental validation in an urban environment with an AV equipped with a modular ROS architecture demonstrated the efficient generation of complex motion paths, ensuring continuity in curvature and heading angle, thereby enhancing the efficiency and stability of path planning \cite{li2020efficent}.

Liao et al. \cite{liao2021game} addressed the challenges posed by ramp merging in traffic, known for its chaotic nature and significant contribution to accidents and congestion. This paper proposes a game theory-based strategy to tackle these issues. By leveraging the collaborative capabilities of connected and automated vehicles during merging maneuvers, the strategy aims to optimize coordination in mixed-traffic environments that include legacy vehicles.\\
The effectiveness of the proposed strategy is demonstrated through simulations conducted on an integrated Unity-SUMO platform. The results show substantial improvements, including up to a 210\% increase in traffic flow speed and a reduction of up to 53.9\% in fuel consumption. Additionally, the strategy helps stabilize driving volatility, ensuring smoother and safer merging operations in mixed traffic scenarios.

Numerous lane-change algorithms have been developed to assist AVs in navigating lane changes across various traffic scenarios. These algorithms utilize a range of methodologies to simulate driver behavior and make lane-change decisions, each designed to address specific simulation requirements and objectives. Each algorithm offers unique features and capabilities \cite{tang2021collision}, allowing researchers and practitioners to model diverse traffic conditions and thoroughly evaluate different aspects of traffic flow dynamics. This versatility enables a comprehensive analysis and optimization of traffic management strategies.

Rooted in the Krauß and MOBIL models, LC2013 is a foundational lane change algorithm in SUMO. It assesses safety gaps and follower benefits to balance safety and traffic flow. Building on LC2013, LC2014 improves decision-making with additional safety criteria, enhancing the handling of various scenarios. LC2017 advances this further by providing more accurate predictions in complex scenarios with multiple followers, improving simulation realism in dense traffic.

The MOBIL model prioritizes traffic efficiency by optimizing lane changes to minimize braking and maximize flow, making it effective in scenarios focused on traffic optimization. The IDM/LC model integrates the Intelligent Driver Model with a lane change algorithm, simulating realistic driver behavior by considering factors like desired time gaps and safety margins. The Continuous Model offers smoother, gradual adjustments in lane position, simulating more natural and realistic lane change behavior compared to discrete changes.

\section{Methodology}

This section delves deeper into the proposed lane-changing and merging strategy tailored for AVs navigating through mixed traffic scenarios and highways. To illustrate, consider a common scenario in which a ramp merges with a highway. Typically, the speed limit on the ramp ranges from $40$km/h to $50$km/h, whereas on a highway, it is approximately $80$ km/h. 

The primary safety-critical challenge lies in effectively executing the ramp merging maneuver. The core idea is to maintain the AV in a specific lane with a safe speed limit for as long as possible, thereby minimizing unnecessary lane changes. For instance, envision a scenario where the AV enters a highway and travels approximately 10 km. Assuming that a highway comprises five lanes, upon entry, the AV merges with the traffic flow and positions itself in the middle lane \cite{liao2021game,li2020efficent}.
\subsection{Scenarios}
\subsubsection{Lane Change Considerations for one AV}

Consider a situation where the left lane is relatively empty, allowing the AV to potentially switch to it and increase its speed by $3$km/h. However, any decision to change lanes triggers the merging algorithm. This introduces potential risks, such as the failure of obstacle detection sensors or adverse weather conditions. In safety-critical domains such as AV navigation, prioritizing safety over speed is paramount. Thus, minimizing lane changes reduces the overall risk of accidents, thereby ensuring the well-being of passengers and other road users.

Highlighting the cascading effects of frequent lane changes and mergers in high-speed environments is essential. Each lane change introduces uncertainty and potential hazards, particularly when interacting with other vehicles traveling at varying speeds. The increased complexity of navigation in such scenarios increases the likelihood of accidents or disruptions in traffic flow \cite{cao2017optimal}. Moreover, frequent lane changes affect the efficiency of traffic flow. Each maneuver requires time and space, contributing to delays and congestion, particularly during peak hours and in densely populated areas. By minimizing unnecessary lane changes, the overall efficiency of the transportation system can be enhanced, leading to a smoother traffic flow and reduced travel times for all road users.

The proposed strategy aims to balance safety and efficiency in AV operations. By prioritizing stable lane positioning and minimizing unnecessary maneuvers, the risk of accidents can be mitigated while optimizing traffic flow. This approach aligns with the broader goals of advancing autonomous driving technologies toward safer and more efficient transportation systems.

\subsubsection{Lane Change Considerations in V2V Communication}

In another scenario, consider the concept of V2V communication in AVs. Building upon the previous scenario, the key difference lies in the presence of multiple interconnected autonomous cars, enabling real-time communication and coordination. However, if one of the vehicles experiences a failure in obstacle detection, suboptimal lane-changing and merging decisions may result. Consider a situation in which an AV fails to detect an obstacle and initiates a lane change. This action, based on flawed information, can potentially disrupt the flow of traffic and create a cascade effect. Another AV intending to change lanes may encounter unexpected obstacles or sudden maneuvers from neighboring vehicles, leading to a chain reaction of lane changes and potential accidents \cite{cao2017optimal}.

\begin{figure}[H]
\centering
\resizebox{1.0\columnwidth}{!}{
\begin{tikzpicture}[auto,node distance=5cm,semithick,
    state/.style={draw,circle,minimum size=2cm},
    point/.style={draw,circle,fill,inner sep=0.0001pt}]
    \node[state,initial] (I) {Idle};
    \node[state] (W) [above right of=I] {Waiting};
    \node[state] (L) [below left of=I] {Moving Left};
    \node[state] (R) [below right of=I] {Moving Right};
    
 \path[->]
    (I) edge [bend left=20] node[above,pos=0.5,at={(0.23,0.4)}] {N \& W} (W)
    (I) edge [bend right=20] node[left,pos=0.5,at={(-0.3,-0.4)}] {N \& L} (L)
    (I) edge [bend left=20] node[right,pos=0.5,at={(0.3,-0.4)}] {N \& R} (R)
    (W) edge [bend left=20] node[below,pos=0.5,at={(0.3,-0.5)}] {N \& (L $\lor$ R)} (I)
    (L) edge [loop left] node[pos=0.5,at={(-0.3,0.0)}] {N \& L} (L)
    (L) edge [bend right=20] node[above,pos=0.5,at={(0.3,0.4)}] {N \& $\lnot$L} (I)
    (R) edge [loop right] node[pos=0.5,at={(0.3,0.0)}] {N \& R} (R)
    (R) edge [bend left=20] node[above,pos=0.5,at={(0.3,0.4)}] {N \& $\lnot$R} (I)
    (I) edge [loop above] node[pos=0.5,at={(0,0.4)}] {$\lnot$N} (I)
    (W) edge [loop above] node[pos=0.5,at={(0,0.4)}] {N \& $\lnot$(L $\lor$ R)} (W);
    
    % Edit points
    \node[point] (E1) at (I.north) {};
    \node[point] (E2) at (W.east) {};
    \node[point] (E3) at (L.west) {};
    \node[point] (E4) at (R.east) {};
\end{tikzpicture}
}
\caption{Finite-state machine diagram for AV lane-change decision logic. \textbf{States:} \textit{Idle} (no active request), \textit{Waiting} (evaluating safety gaps), \textit{Moving Left/Right} (executing lane-change maneuver). \textbf{Boolean triggers:} $N$ (lane-change needed), $W$ (waiting timeout), $L$/$R$ (safe gap on left/right). Solid arrows indicate conditional transitions; loops represent state persistence or fallback logic based on gap availability or timeout.}
\label{fig:StateMachineDiagram}
\end{figure}
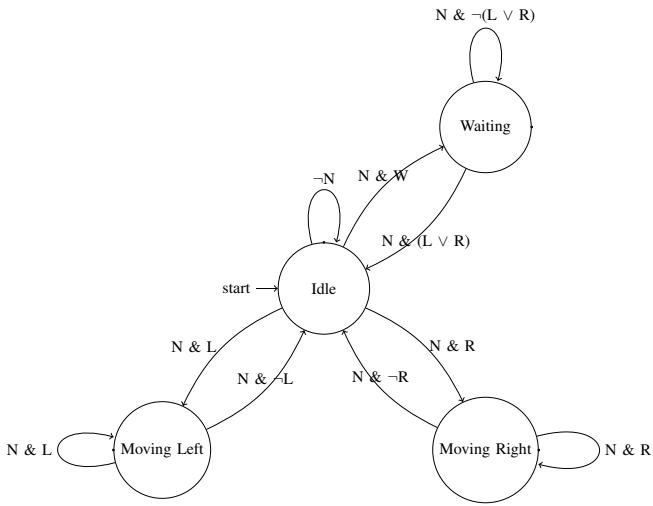

Therefore, adhering to maintaining a safe lane becomes even more crucial in V2V scenarios. Minimizing unnecessary lane changes can significantly reduce the risk of such cascading incidents. Rather than prioritizing rapid lane changes for individual vehicles, the focus shifts toward collective safety and stability within traffic flow.

Furthermore, the concept of V2V communication opens avenues for proactive risk mitigation strategies. For example, if an AV detects an obstacle or hazardous condition, it can promptly relay this information to nearby vehicles, thereby enabling them to adjust their trajectories accordingly. This collaborative approach enhances situational awareness and promotes safer decision-making among all vehicles in the network. While V2V communication offers tremendous potential for enhancing traffic safety and efficiency, it also introduces new challenges related to coordination and reliability. By emphasizing the importance of staying in a safe lane and fostering cooperative behavior among AVs, we can mitigate these challenges and pave the way for safer and more reliable autonomous driving ecosystems.

\subsection{ MLCA Algorithm}

An algorithmic workflow can significantly improve the safety of AVs by offering a structured approach to their operations. MLCA Algorithm is a roadmap, empowering AVs to navigate roads with enhanced safety measures. Although MLCA presents a fundamental algorithm designed for conceptual understanding within a contextual information setting rather than a practical application, it lays the groundwork for illustrating its real-time implementation within an experimental plan section.

\begin{algorithm}
\caption{MLCA Algorithm for Autonomous Vehicle Movement Control}
\label{AVStateMachine}
\begin{algorithmic}[1]
\REQUIRE Navigation signal: \texttt{N}, Wait command: \texttt{W}, Left command: \texttt{L}, Right command: \texttt{R}
\ENSURE Current state of the vehicle: \texttt{state}
\STATE \textbf{Initialize:} \texttt{current\_state} $\gets$ IDLE
\STATE \textbf{Define States:} IDLE, WAITING, MOVING\_LEFT, MOVING\_RIGHT

\WHILE{system is active}
    \IF{$\neg$\texttt{N}}
        \STATE \texttt{current\_state} $\gets$ IDLE
        \STATE \textbf{continue}
    \ENDIF

    \IF{\texttt{current\_state} = IDLE}
        \IF{\texttt{N} $\land$ \texttt{W}}
            \STATE \texttt{current\_state} $\gets$ WAITING
        \ELSIF{\texttt{N} $\land$ \texttt{L}}
            \STATE \texttt{current\_state} $\gets$ MOVING\_LEFT
        \ELSIF{\texttt{N} $\land$ \texttt{R}}
            \STATE \texttt{current\_state} $\gets$ MOVING\_RIGHT
        \ENDIF
    
    \ELSIF{\texttt{current\_state} = WAITING}
        \IF{\texttt{N} $\land$ (\texttt{L} $\lor$ \texttt{R})}
            \STATE \texttt{current\_state} $\gets$ IDLE
        \ENDIF
    
    \ELSIF{\texttt{current\_state} = MOVING\_LEFT}
        \IF{\texttt{N} $\land$ $\neg$\texttt{L}}
            \STATE \texttt{current\_state} $\gets$ IDLE
        \ENDIF
    
    \ELSIF{\texttt{current\_state} = MOVING\_RIGHT}
        \IF{\texttt{N} $\land$ $\neg$\texttt{R}}
            \STATE \texttt{current\_state} $\gets$ IDLE
        \ENDIF
    \ENDIF

    \STATE \textbf{Assert:} \texttt{current\_state} $\in$ \{IDLE, WAITING, MOVING\_LEFT, MOVING\_RIGHT\}
    \STATE \textbf{Assert:} $\neg$\texttt{N} $\implies$ \texttt{current\_state} = IDLE
    \STATE \textbf{Assert:} \texttt{current\_state} = MOVING\_LEFT $\implies$ (\texttt{N} $\land$ \texttt{L})
    \STATE \textbf{Assert:} \texttt{current\_state} = MOVING\_RIGHT $\implies$ (\texttt{N} $\land$ \texttt{R})
    \STATE \textbf{Output:} \texttt{state} $\gets$ \texttt{current\_state}
\ENDWHILE
\end{algorithmic}
\end{algorithm}

The MLCA algorithm, as outlined in the AV movement decision state machine in Fig. \ref{fig:StateMachineDiagram} and detailed in Alg. \ref{AVStateMachine}, is specifically designed to regulate the movement behavior of an autonomous vehicle (AV) in a multi-lane environment. The algorithm defines four distinct operational states: Idle, Waiting, Moving Left, and Moving Right. The transitions between states are dictated by a set of Boolean variables: N (Need to Move), W (Can Wait), L (Left Side Empty), and R (Right Side Empty). The AV begins in the Idle state, where it remains stationary. A transition occurs from the Idle state to the Waiting state if there is a need to move AND the AV can wait (N AND W). Similarly, the AV transitions to the Moving Left or Moving Right states if movement is necessary AND the respective adjacent lane is clear (N AND L or N AND R). Upon entering the Waiting state, the AV monitors lane conditions, allowing it to revert to the Idle state when movement is no longer required (NOT N AND (L OR R)). The Moving Left and Moving Right states are maintained through self-loops as long as the need to move AND the corresponding lane condition persist. If the conditions for these states change, such as the need to stop or an obstacle appearing, the AV returns to the Idle state (N AND NOT L or N AND NOT R). Self-loops are incorporated to maintain the current state when conditions remain unchanged. Edit points, represented by filled circles on each state, are included to facilitate potential modifications or extensions to the state machine's behavior. This design ensures a robust and adaptable decision-making process for AV movement in various traffic scenarios.

In addition to addressing collision avoidance in these scenarios, various strategies and algorithms proposed in academic papers provide guidelines for mitigating the risk of collisions \cite{funke2016collision}. Therefore, efforts were made to implement the algorithm within a controlled testing environment to conduct thorough assessments and compare its performance with and without implementation. This approach involved meticulously designing the test environment, ensuring controlled variables for precise comparison and thoroughly evaluating the algorithm's efficacy.

\section{Experimental Setup and Analysis}
In safety-critical embedded systems, the primary concern is always ensuring safety. Despite AVs' state-of-the-art technologies and sophisticated machine learning algorithms, the fundamental question remains: ``Is it safe?" The inherent risks associated with AVs demand meticulous risk mitigation strategies to safeguard passengers and other vehicles sharing the road. One strategy involves minimizing lane changes. By reducing the frequency of lane changes, potential risks can be effectively mitigated, enhancing overall safety.

Each lane change introduces uncertainty and vulnerability because it requires the AV to navigate dynamic traffic conditions and interact with other vehicles. In addition, factors such as sensor limitations, unpredictable behavior of other drivers, and environmental variables further compound the risks associated with lane changes. Therefore, prioritizing stability and consistency in lane positioning can significantly reduce risk. By maintaining a steady trajectory within a designated lane, AVs can minimize their exposure to potential hazards and mitigate the likelihood of accidents.

Furthermore, limiting lane changes promotes smoother traffic flow and reduces the potential for disruptions or conflicts with other vehicles on the road. While AVs showcase cutting-edge technologies and capabilities, ensuring their safety remains paramount. By adopting a proactive approach to risk management, such as minimizing lane changes, the safety of AV passengers and other road users can be enhanced, ultimately advancing the acceptance and integration of autonomous driving technologies into our transportation systems.

To rigorously evaluate the effectiveness of this risk mitigation strategy, a comprehensive testing environment was created using SUMO and OpenStreetMap to simulate various scenarios and assess the algorithm's performance. By incorporating these advanced simulation tools, real-world conditions can be closely mimicked, thoroughly assessing the algorithm's behavior across different scenarios. Through meticulous testing and comparison, valuable insights into the algorithm's efficacy and potential areas for improvement were obtained. This approach ensures that the proposed strategy not only enhances safety theoretically but also demonstrates practical effectiveness in diverse and dynamic traffic conditions.

\subsection{Tools and Environment Setup}
\subsubsection{OpenStreetMap} Provides free geographic data, including roads, buildings, and other infrastructure. These open datasets were imported into SUMO and CARLA to recreate real-world landscapes for testing.

\subsubsection[Simulation of Urban Mobility]{Simulation of Urban Mobility \cite{krajzewicz2010traffic}}
SUMO is an open-source, highly portable, microscopic traffic simulation package. This allows the importation of OpenStreetMap data and the simulation of real-world traffic flows. SUMO enables the configuration of vehicles, routes, and mobility models, providing a robust platform for simulating traffic scenarios.

\subsubsection{Hardware Configuration}
All simulations were performed sequentially on a Windows 11 workstation powered by an Intel Core i5‑8250U processor (1.60 GHz, 4 cores/8 threads) with 16 GB of RAM.

\subsection{Simulation Scenarios}
SUMO was used to test various scenarios with a small population of cars ($100$ cars on the road)  before and after implementing the proposed algorithm over a total distance of $20$ km. This approach facilitates the evaluation of the effectiveness and practical utility of the algorithm. By conducting these simulations, the goal was to quantify the improvements brought about by the algorithm and assess its value in optimizing the traffic flow and enhancing overall road safety.

\subsection{Simulation Setup} The first step involved setting up the simulation environment based on the scenarios provided. SUMO utilizes XML to define simulation scenarios, network configurations, and various parameters. To align the simulation to minimize lane changes, the attributes for each vehicle and road configuration were customized \cite{sumo}. This ensured that the simulation was appropriately tailored to meet specific objectives.

In the simulation, several key attributes were incorporated to enhance realism, including:
\begin{itemize}
    \item \texttt{departLane}: Specifies the lane on which the vehicle should be inserted into the network.
   \item \texttt{departPos}: Defines the position at which the vehicle will enter the network.
    \item \texttt{departEdge}: Determines the initial edge along the route where the vehicle enters the network.
    \item \texttt{arrivalLane}: Indicates the lane at which the vehicle will exit the network.
    \item \texttt{arrivalPos}: Specifies the position at which the vehicle will leave the network.
    \item \texttt{arrivalEdge}: Specifies the final edge along the route where the vehicle exits the network.
    \item \texttt{departPosLat}: Specifies the lateral position on the departure lane at which the vehicle enters the network.
\end{itemize}

These attributes, along with other elements and configurations, were carefully selected and implemented to create a more realistic simulation environment.

\subsection{Simulation Execution and Data Collection}
Simulation scenarios A and B were executed sequentially in SUMO for 100 iterations, during which the green vehicle designated as the primary data source was monitored (vehicle color assignments were configured within the SUMO environment). This iterative procedure yielded a comprehensive dataset for assessing MLCA’s impact. The complete set of simulations ran on a Windows 11 workstation equipped with an Intel Core i5‑8250U (1.60 GHz, 4 cores/8 threads) and 16 GB of RAM, requiring approximately six hours and 30 minutes of wall‑clock time, with average CPU utilization of 75\% and peak memory usage of 6GB.

Post-simulation graphs were generated to visualize the results obtained from each scenario. These visualizations facilitated an understanding of the algorithm's effectiveness in minimizing lane changes and enhancing overall traffic flow. Improvements in safety and efficiency were quantified by comparing pre-and post-implementation data, providing valuable insights into the algorithm's performance.

This detailed experimental plan aimed to rigorously test and validate the proposed lane-changing and merging strategy for AVs. The insights gained from these simulations can contribute to developing safer and more efficient autonomous driving technologies.

\subsection{Results}

The analysis of lane change data was conducted using three distinct charts, each representing different model algorithms over a cumulative distance of 20 kilometers.

\subsubsection{LC2017 and MOBIL Algorithms}

\begin{figure}[htbp]
    \centering
    \resizebox{1.0\columnwidth}{!}{ 
    \begin{tikzpicture}
    \begin{axis}[
        ybar,
        bar width=22pt,
        width=\textwidth,
        height=0.45\textwidth,
        legend style={at={(0.23,0.8)}, anchor=south, legend columns=2, font=\large},
        ylabel={Number of Lane Changes},
        ylabel style={font=\large},
        xlabel={Cumulative Distance (KM)},
        xlabel style={at={(0.5,-0.15)}, font=\Large},
        symbolic x coords={5,10,15,20},
        xtick=data,
        xticklabels={5 KM ,10 KM,15 KM,20 KM},
        x tick label style={font=\Large, text width=3.5cm, align=center}, % Increased font size for x-axis labels
        nodes near coords,
        every node near coord/.append style={font=\large}, % Increased font size for numbers
        nodes near coords align={vertical},
        ymin=0, ymax=15,
        title={Number of Lane changes using LC2017 algorithm and MOBIL algorithm},
        title style={at={(0.5,1.0)},font=\large},
        tick label style={font=\large},
        label style={font=\large},
        enlarge x limits=0.2, % Increased space on left and right sides
        ylabel near ticks, % Adds space between y-axis numbers and label
        y tick label style={/pgf/number format/.cd, fixed, fixed zerofill, precision=0, /tikz/.cd}, % Removes decimal points from y-axis labels
    ]
    \addplot[pattern=north east lines, pattern color=olive!60] coordinates {
        (5,1) (10,2) (15,4) (20,5)
    };
    \addplot[pattern=crosshatch, pattern color=blue!60] coordinates {
        (5,1) (10,2) (15,3) (20,4)
    };
    \addplot[pattern=grid, pattern color=black!60] coordinates {
        (5,2) (10,6) (15,9) (20,14)
    };
    \addplot[pattern=crosshatch dots, pattern color=red!60] coordinates {
        (5,2) (10,5) (15,9) (20,13)
    };
    \legend{LC2017 (1 AV), MOBIL (1 AV), LC2017 (3 AVs), MOBIL. (3 AVs)}
    \end{axis}
    \end{tikzpicture}}
    \caption{Comparison of the Number of Lane Changes at 5, 10, 15, and 20 Kilometers Using LC2017 and MOBIL Algorithms Over 100 Iterations}
    \label{fig:LC2017}
\end{figure}
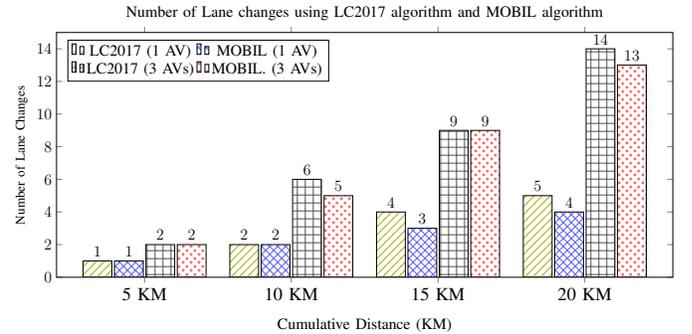

Based on demonstrated sample outcomes in Fig. [\ref{fig:LC2017}], in one AV Scenario, the number of lane changes in the LC2017 Model is initially low and gradually increases, reaching approximately \textbf{five} lane changes by the 20-kilometer mark. However, in the MOBIL Model Algorithm, the number of lane changes peaking around \textbf{four} lane changes at 20 kilometers.

In addition, in the Three AVs Scenario, more lane changes are observed in the LC2017 Model Algorithm compared to the one AV scenario, reaching around \textbf{fourteen} lane changes by 20 km. The MOBIL Model Algorithm consistently increases in lane changes, reaching approximately \textbf{thirteen} lane changes by 20 km.

\subsubsection{ IDM/LC and Continuous Algorithms}

\begin{figure}[htbp]
    \centering
    \resizebox{1.0\columnwidth}{!}{ 
    \begin{tikzpicture}
    \begin{axis}[
        ybar,
        bar width=22pt,
        width=\textwidth,
        height=0.45\textwidth,
        legend style={at={(0.25,0.8)}, anchor=south, legend columns=2, font=\large},
        ylabel={Average Number of Lane Changes},
        ylabel style={font=\large},
        xlabel={Cumulative Distance (KM)},
        xlabel style={at={(0.5,-0.15)}, font=\Large},
        symbolic x coords={5,10,15,20},
        xtick=data,
        xticklabels={5 KM ,10 KM,15 KM,20 KM},
        x tick label style={font=\Large, text width=3.5cm, align=center},
        nodes near coords,
        every node near coord/.append style={font=\large},
        nodes near coords align={vertical},
        ymin=0, ymax=15,
        title={Number of Lane changes using IDM/LC algorithm and Continuous algorithm},
        title style={at={(0.5,1.0)},font=\large},
        tick label style={font=\large},
        label style={font=\large},
        enlarge x limits=0.2,
        ylabel near ticks,
        y tick label style={/pgf/number format/.cd, fixed, fixed zerofill, precision=0, /tikz/.cd},
    ]
    \addplot[pattern=north east lines, pattern color=olive!60] coordinates {
        (5,1) (10,2) (15,4) (20,6)
    };
    \addplot[pattern=crosshatch, pattern color=blue!60] coordinates {
        (5,0) (10,1) (15,3) (20,4)
    };
    \addplot[pattern=grid, pattern color=black!60] coordinates {
        (5,3) (10,7) (15,8) (20,13)
    };
    \addplot[pattern=crosshatch dots, pattern color=red!60] coordinates {
        (5,3) (10,6) (15,9) (20,12)
    };
    \legend{IDM/LC (1 AV), Continuous (1 AV), IDM/LC (3 AVs), Continuous (3 AVs)}
    \end{axis}
    \end{tikzpicture}}
    \label{fig:IDM}
\caption{Comparison of Lane Changes at 5, 10, 15, and 20 Kilometers Using the IDM/LC and Continuous Algorithms Over 100 Iterations}
\end{figure}
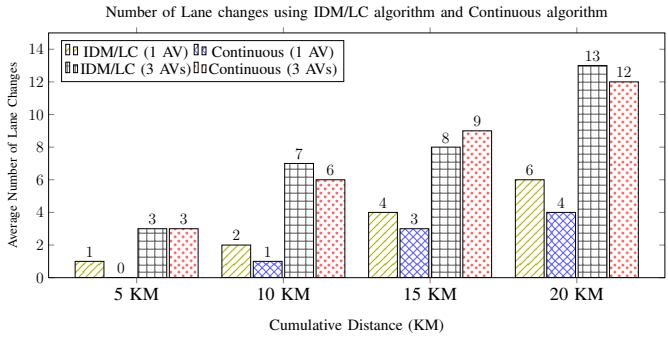

The IDM/LC model algorithm exhibited low initial lane changes based on depicted sample results in Fig. [\ref{fig:IDM}] , slightly increasing to about \textbf{six} changes at 20 KM in the one AV scenario. For three AVs, lane changes increased more noticeably, peaking at around \textbf{thirteen}. This indicates that the IDM/LC model promotes cautious driving behavior but adapts to optimize traffic flow with more AVs.

The Continuous model algorithm showed a steady rise in lane changes. The one AV scenario reached approximately \textbf{four} changes by 20 KM, while the three AV scenarios indicated a higher frequency of lane changes, peaking at around \textbf{twelve}. This pattern suggests that the Continuous model balances and adapts lane-changing behavior based on traffic conditions and AV presence.

\subsubsection{ MLCA and Without Applied Algorithm}

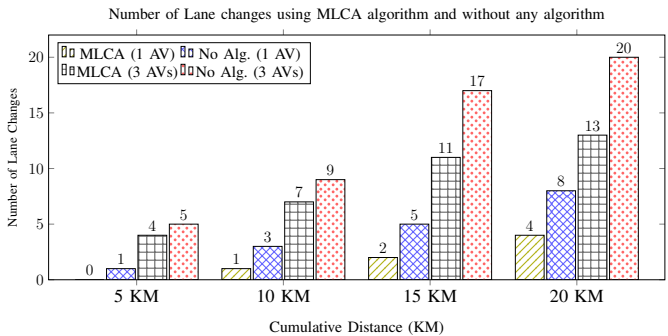
\begin{figure}[htbp]
    \centering
    \resizebox{1.0\columnwidth}{!}{ % Adjust the width to fit the column
    \begin{tikzpicture}
    \begin{axis}[
        ybar,
        bar width=22pt, % Adjusted bar width for better spacing
        width=1.0\textwidth,
        height=0.45\textwidth,
        legend style={at={(0.22,0.80)}, anchor=south, legend columns=2, font=\large},
        ylabel={Number of Lane Changes},
        ylabel style={font=\large},
        xlabel={Cumulative Distance (KM)},
        xlabel style={at={(0.5,-0.15)}, font=\Large},
        symbolic x coords={5,10,15,20},
        xtick=data,
        xticklabels={5 KM ,10 KM,15 KM,20 KM},
        x tick label style={font=\Large, text width=3.5cm, align=center}, % Increased font size for x-axis labels
        nodes near coords,
        every node near coord/.append style={font=\large}, % Increased font size for numbers
        nodes near coords align={vertical},
        ymin=0, ymax=22,
        title={Number of Lane changes using MLCA algorithm and without any algorithm},
        title style={at={(0.5,1.0)},font=\large},
        tick label style={font=\large},
        label style={font=\large},
        enlarge x limits=0.2, % Increased space on left and right sides
        ylabel near ticks, % Adds space between y-axis numbers and label
        y tick label style={/pgf/number format/.cd, fixed, fixed zerofill, precision=0, /tikz/.cd}, % Removes decimal points from y-axis labels
    ]
    \addplot[pattern=north east lines, pattern color=olive!60] coordinates {

        (5,0) (10,1) (15,2) (20,4)
    };
    \addplot[pattern=crosshatch, pattern color=blue!60] coordinates {
        (5,1) (10,3) (15,5) (20,8)
    };
    \addplot[pattern=grid, pattern color=black!60] coordinates {
        (5,4) (10,7) (15,11) (20,13)
    };
    \addplot[pattern=crosshatch dots, pattern color=red!60] coordinates {
        (5,5) (10,9) (15,17) (20,20)
    };
    \legend{MLCA (1 AV), No Alg. (1 AV), MLCA (3 AVs), No Alg. (3 AVs)}
    \end{axis}
    \end{tikzpicture}}
\caption{Comparison of the Number of Lane changes at 5, 10, 15, and 20 kilometers Using MLCA algorithm and without applying algorithm Over 100 Iterations}
    \label{fig:my}
\end{figure}

The sample results shown in Fig. [\ref{fig:my}] demonstrate that the MLCA Algorithm exhibited a significant difference in performance between one AV and three AVs. The one AV scenario reached about \textbf{four} lane changes, whereas the three AV scenarios peaked at approximately \textbf{thirteen}. This indicates that MLCA effectively leverages multiple AVs for optimized lane-changing decisions.

In the absence of any applied algorithm, lane changes were minimal in the one AV scenario, reaching about \textbf{eight} changes at 20 KM. The three AVs scenario, however, demonstrated the highest number of lane changes among all scenarios, peaking at approximately \textbf{twenty}. This highlights the potential impact of algorithmic control on traffic dynamics and the necessity of applied algorithms for efficient traffic management.

% \begin{figure}[h!]
%     \centering
%     \includegraphics[width=.75\linewidth]{Picture1.png}
%     \caption{Number of Lane Changes: MLCA vs. Without Algorithm}
%     \label{fig:my}
% \end{figure}

\subsubsection{Comparative Analysis of Collision}

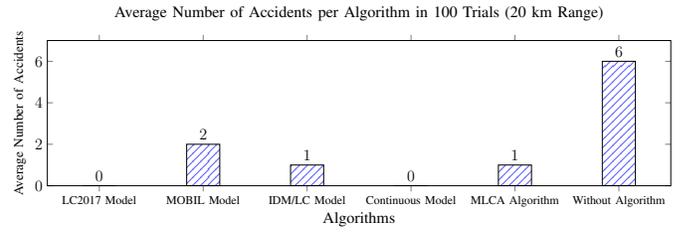
\begin{figure}[htbp]
    \centering
    \resizebox{1.0\columnwidth}{!}{ 
    \begin{tikzpicture}
    \begin{axis}[
        ybar,
        bar width=25pt,
        width=\textwidth,
        height=0.3\textwidth,
        ylabel={Average Number of Accidents},
        ylabel style={font=\large},
        xlabel={Algorithms},
        xlabel style={at={(0.5,-0.15)}, font=\large},
        symbolic x coords={LC2017,MOBIL,IDM/LC,Continuous,MLCA,Without},
        xtick=data,
        xticklabels={LC2017 Model,MOBIL Model,IDM/LC Model,Continuous Model,MLCA Algorithm,Without Algorithm},
        x tick label style={font=\small},
        nodes near coords,
        every node near coord/.append style={font=\large},
        nodes near coords align={vertical},
        ymin=0, ymax=7,
        title={Average Number of Accidents per Algorithm in 100 Trials (20 km Range)},
        title style={at={(0.5,1.05)},font=\large, align=center},
        tick label style={font=\large},
        enlarge x limits=0.1,
        ylabel near ticks,
        y tick label style={/pgf/number format/.cd, fixed, fixed zerofill, precision=0, /tikz/.cd},
    ]
    \addplot[fill=blue!60, pattern=north east lines, pattern color=blue!60] coordinates {
        (LC2017,0) (MOBIL,2) (IDM/LC,1) (Continuous,0) (MLCA,1) (Without,6)
    };
    \end{axis}
    \end{tikzpicture}}
    \caption{Comparison of average number of accidents for different algorithms over 100 trials in a 20 km range}
    \label{fig:compare}
\end{figure}

The Fig. [\ref{fig:compare}], which shows the average number of accidents per algorithm over 100 trials within a 20 km range, highlights significant performance differences. Without an algorithm, collisions average about \textbf{seven} per 100 trials, underscoring the importance of algorithmic intervention for vehicular safety. Among the tested algorithms, MLCA has the lowest average number of collisions, proving its superior effectiveness. The IDM/LC Model also performs well, with the second-lowest collision average. The MOBIL Model and LC2017 Model show moderate performance with slightly higher collision averages. The Continuous Model's performance is either not provided or negligible. Implementing algorithms significantly enhances safety, with MLCA reducing accidents the most.

\section{Limitations and Future Work} 

This study focused on minimizing lane changes without addressing collision avoidance, time considerations, or travel duration. Future work should incorporate these aspects and explore alternative scenarios for a more comprehensive analysis.

The study's limitations include the lack of high-level simulator testing. Future research will leverage CARLA, an open-source simulator for autonomous driving, to create realistic driving scenarios for testing and refining perception and control algorithms. Integrating the proposed algorithm with advanced control systems is expected to enhance its effectiveness, while CARLA's collaborative features may foster innovative approaches.

Future work will also explore cybersecurity in V2V communication to ensure secure, reliable connections that enhance safety and reduce lane changes. Addressing cybersecurity vulnerabilities is essential for developing robust systems that are resistant to threats.
Additionally, the research will examine the algorithm's performance under diverse environmental conditions, including adverse weather, varied lighting, and complex traffic patterns, to evaluate its robustness and adaptability.

Incorporating insights from these studies and developing robust techniques for secure V2V communication will contribute significantly to creating dependable and safe autonomous systems. This comprehensive approach will expand expertise and prepare professionals to tackle emerging challenges and seize new opportunities in the rapidly evolving field of autonomous driving technology. The integration of these advancements will pave the way for more resilient and intelligent autonomous systems that can operate effectively in real-world conditions. Future work will also explore the use of distributed computing and GPU-accelerated frameworks, such as MPI clusters or Flow on Ray, to significantly reduce simulation times below the 7 hours observed on a standard laptop.

\section*{Conclusion}
The analysis of the charts and data from the simulations has indicated that the proposed algorithm substantially improves the stability and efficiency of AVs. Specifically, the algorithm achieved a reduction in lane changes by approximately 50\% in the first scenario and 33\% in the second scenario. This significant decrease in lane changes not only enhances traffic flow but also lowers the likelihood of collisions, thereby improving overall road safety. Additionally, the reduction in collisions further confirms the algorithm's effectiveness in enhancing AV safety within complex traffic environments.

Despite these promising results, there are considerable opportunities for further refinement. Future research will focus on integrating the algorithm with more advanced control systems and testing it in a broader range of realistic scenarios using platforms like CARLA. In summary, while the proposed algorithm has demonstrated significant improvements in simulated environments, ongoing advancements and extensive testing are crucial to validate its effectiveness and reliability for real-world applications. Addressing these challenges will be key to advancing safer and more efficient autonomous driving technology. On a standard laptop (Core i5‑8250U), MLCA’s full evaluation required 6 hours and 30 minutes over 100 runs, underscoring the benefit of exploring parallel and GPU‑based implementations in future studies.

\bibliography{Reference}

\begin{thebibliography}{10}

\bibitem{li2017development}
X.~Li, Z.~Sun, D.~Cao, D.~Liu, and H.~He, ``Development of a new integrated local trajectory planning and tracking control framework for autonomous ground vehicles,'' {\em Mechanical Systems and Signal Processing}, vol.~87, pp.~118--137, 2017.

\bibitem{li2015practical}
X.~Li, Z.~Sun, Z.~He, Q.~Zhu, and D.~Liu, ``A practical trajectory planning framework for autonomous ground vehicles driving in urban environments,'' in {\em 2015 IEEE Intelligent Vehicles Symposium (IV)}, pp.~1160--1166, IEEE, 2015.

\bibitem{hubmann2017decision}
C.~Hubmann, M.~Becker, D.~Althoff, D.~Lenz, and C.~Stiller, ``Decision making for autonomous driving considering interaction and uncertain prediction of surrounding vehicles,'' in {\em 2017 IEEE intelligent vehicles symposium (IV)}, pp.~1671--1678, IEEE, 2017.

\bibitem{AV}
M.~Placek, ``Autonomous vehicles worldwide – statistics \& facts.'' [Online]. Available: \url{https://www.statista.com/topics/3573/autonomous-vehicle-technology/#topicOverview}.
\newblock [Accessed: Dec. 10, 2023].

\bibitem{liu2023vehicle}
X.~Liu, L.~Hong, and Y.~Lin, ``Vehicle lane change models—a historical review,'' {\em Applied Sciences}, vol.~13, no.~22, p.~12366, 2023.

\bibitem{kesting2007general}
A.~Kesting, M.~Treiber, and D.~Helbing, ``General lane-changing model mobil for car-following models,'' {\em Transportation Research Record}, vol.~1999, no.~1, pp.~86--94, 2007.

\bibitem{chamieh2012impact}
M.~Chamieh, R.~El-Kouatly, and O.~Abu-Amsha, ``Impact of idm lane-changing on the performance of aodv on vanets,'' {\em International Journal of Simulation Systems, Science \& Technology (IJSSST)}, vol.~13, no.~6, 2012.

\bibitem{wang2019continuous}
P.~Wang, H.~Li, and C.-Y. Chan, ``Continuous control for automated lane change behavior based on deep deterministic policy gradient algorithm,'' in {\em 2019 IEEE Intelligent Vehicles Symposium (IV)}, pp.~1454--1460, IEEE, 2019.

\bibitem{li2020efficent}
Z.~Li, H.~Liang, P.~Zhao, S.~Wang, and H.~Zhu, ``Efficent lane change path planning based on quintic spline for autonomous vehicles,'' in {\em 2020 IEEE International Conference on Mechatronics and Automation (ICMA)}, pp.~338--344, IEEE, 2020.

\bibitem{liao2021game}
X.~Liao, X.~Zhao, Z.~Wang, K.~Han, P.~Tiwari, M.~J. Barth, and G.~Wu, ``Game theory-based ramp merging for mixed traffic with unity-sumo co-simulation,'' {\em IEEE Transactions on Systems, Man, and Cybernetics: Systems}, vol.~52, no.~9, pp.~5746--5757, 2021.

\bibitem{tang2021collision}
Y.~Tang, Y.~Zhou, Y.~Liu, J.~Sun, and G.~Wang, ``Collision avoidance testing for autonomous driving systems on complete maps,'' in {\em 2021 IEEE Intelligent Vehicles Symposium (IV)}, pp.~179--185, IEEE, 2021.

\bibitem{cao2017optimal}
P.~Cao, Y.~Hu, T.~Miwa, Y.~Wakita, T.~Morikawa, and X.~Liu, ``An optimal mandatory lane change decision model for autonomous vehicles in urban arterials,'' {\em Journal of Intelligent Transportation Systems}, vol.~21, no.~4, pp.~271--284, 2017.

\bibitem{funke2016collision}
J.~Funke, M.~Brown, S.~M. Erlien, and J.~C. Gerdes, ``Collision avoidance and stabilization for autonomous vehicles in emergency scenarios,'' {\em IEEE Transactions on Control Systems Technology}, vol.~25, no.~4, pp.~1204--1216, 2016.

\bibitem{krajzewicz2010traffic}
D.~Krajzewicz, ``Traffic simulation with sumo--simulation of urban mobility,'' {\em Fundamentals of traffic simulation}, pp.~269--293, 2010.

\bibitem{sumo}
S.~S. of~Urban~MObility, ``Definition of vehicles, vehicle types, and routes.''

\end{thebibliography}
\bibliographystyle{ieeetr}

\end{document}